\documentclass[preprint,showpacs,preprintnumbers,amsmath,amssymb,nofootinbib]{revtex4}
\usepackage{epsfig}

\begin{document}

\title{Antibound States and Halo Formation in the Gamow Shell Model}

\author{N.~Michel$^{1-3}$}
\author{W.~Nazarewicz$^{1,2,4}$}
 \author{M.~P{\l}oszajczak$^{5}$}
 \author{J.~Rotureau$^{1-3}$}

\affiliation{%
$^1$Department of Physics and Astronomy, University of Tennessee,
Knoxville, TN 37996 \\
$^2$Physics Division, Oak Ridge National Laboratory, P.O.~Box 2008,
Oak Ridge, TN 37831 \\
$^3$Joint Institute for Heavy-Ion Research, Oak Ridge, TN 37831 \\
$^4$Institute of Theoretical Physics, Warsaw University, ul.~Ho\.{z}a
69, 00-681 Warsaw, Poland \\
$^5$Grand Acc\'{e}l\'{e}rateur National d'Ions Lourds (GANIL),CEA/DSM -- 
CNRS/IN2P3, BP 55027, F-14076 Caen Cedex 05, France
}

\begin{abstract}
The open quantum system formulation of the nuclear shell model, the
so-called Gamow Shell Model (GSM), is a multi-configurational SM  
that employs a
single-particle basis given by the Berggren ensemble consisting
of Gamow  states and the  non-resonant continuum of scattering
states.  
The GSM is of particular importance for
 weakly bound/unbound nuclear states where both
many-body correlations and the coupling to 
decay channels are essential.
In this context, we investigate
the role of $\ell$=0 antibound (virtual) neutron single-particle states in
 the shell model
description of loosely bound wave functions, such
as the ground state  wave function of a halo nucleus $^{11}$Li. 
\end{abstract}
\maketitle

\section{Introduction}
The theoretical description of strongly correlated open quantum systems
(OQS), such as the weakly bound/unbound atomic nuclei or atomic
clusters, requires the rigorous treatment of both the many-body
correlations and the continuum of positive-energy scattering states and decay
channels \cite{(Dob98),(Oko03)}.  A major theoretical challenge is to
consistently describe many-body states close to particle-emission
thresholds where novel properties appear, such as, e.g.,
unusual radial features of halo states or threshold anomalies in the
wave functions and associated observables.  
These features cannot be
described in the closed quantum system (CQS) framework of a standard
shell model (SM) which usually employs the single-particle (s.p.) basis of
${\cal L}^2$-functions of the harmonic oscillator. Representation of
halo states in such a basis is not convenient.
Moreover, the resonance and scattering states do not belong to the space of ${\cal
L}^2$-functions.

The solution of the configuration interaction problem 
in the presence of continuum states (the so-called Continuum Shell Model)
has been advanced recently in the OQS
formulation of the nuclear SM, such as the Shell Model Embedded in the
Continuum \cite{karim,(Rot06)}
(real-energy continuum SM)  and the Gamow Shell Model (GSM)
\cite{(Mic02),(Mic03),(Bet02),(Bet03),(Mic04v1),(Rot06a),(Bet04),(Bet05),(Hag05),(Hag06)} 
(SM in the complex $k$-plane). A  variant of the real-energy
continuum SM, based on the Feshbach projection formalism 
and phenomenological continuum-coupling,
  has been proposed in Ref.~\cite{(Vol05)}.

In the GSM, the multi-configurational SM is formulated in the rigged
Hilbert space \cite{(Rig1),(Rig2)} with a  s.p. basis
given by the Berggren ensemble \cite{(Ber68),(Ber93)} consisting of
Gamow  (resonant or Siegert)  states and the  non-resonant
continuum of scattering states. In general, 
resonant states correspond to the poles
of the scattering matrix ($S$-matrix). These  are the generalized
eigenstates of the time-independent Schr\"{o}dinger equation, which are
regular at the origin and satisfy purely outgoing boundary conditions.
The s.p.~Berggren basis is generated by a finite-depth potential, and
the many-body states are expanded in Slater determinants spanned by
resonant and non-resonant s.p.~basis states \cite{(Mic02),(Bet02)}.  The
configuration mixing induced by the GSM Hamiltonian 
assures a simultaneous  treatment of continuum effects
and inter-nucleon correlations. Consequently,
the GSM, which is a natural generalization of the SM for OQS, gives a
natural description of many-body loosely bound and resonant states.
In general, among different poles of the one-body $S$-matrix, one takes
into account  bound and decaying  s.p.~states  to define the subset of
resonant states in the Berggren ensemble. These poles, together  with
associated non-resonant continuum states,  define the valence space. The
actual selection of resonant states depends on the physical problem and
on the convergence properties of resulting  many-body GSM states.

An important question concerning the GSM deals with the inclusion of
antibound (virtual) states in the Berggren ensemble. 
Antibound states have real and negative energy eigenvalues
that are located in the second Riemann sheet of the complex energy plane
(the corresponding momentum lies on the negative imaginary axis)
\cite{(New66),(Nus72),(Tay72),(Dom81)}. Contrary to bound states,
the radial wave functions of virtual states  increase
exponentially at large distances.  As often discussed in the literature, it is difficult
to give a clear physical interpretation to virtual states. 
Strictly  speaking, as the second
energy sheet is considered unphysical and unaccessible through direct 
experiments,
a virtual state is  not a  state but
a feature of the system.
If  the virtual state has a
sufficiently small  energy, its presence has an appreciable influence on
 the  behavior of the scattering cross section at low energies.
Classic examples include   the low-energy 
$1S_0$ nucleon-nucleon scattering characterized by a large and
negative scattering length \cite{(Tay72)},  scattering of slow electrons on molecules
\cite{(Dub77),(Mor82),(Mor92)}, and $eep$ Coulomb system \cite{(Tol97)}.
Related to this is an increased localization of real-energy scattering
states just above threshold \cite{(Mig71)}. 

Coming back to nuclear structure, it was argued
that the neutron-unbound $^{10}$Li nucleus sustains a low-lying
$1s_{1/2}$ antibound state very close to the one-neutron (1n) emission
threshold  \cite{(Tho94)} as a result of the inversion of $0p_{1/2}$ and
$1s_{1/2}$  shells \cite{(Bro98)}. Although many theoretical
calculations predict the $0p_{1/2} - 1s_{1/2}$ shell inversion,  the phenomenon
still remains a matter of debate \cite{Wur96,Des97}.  Experimentally, several
groups have reported  evidence  of the 
$\ell=0$ strength at the 1n-threshold
in $^{10}$Li  \cite{Shi97,Zin97,Shi98,You94,Kob93,(Zin95)}; however, no
evidence of  a weakly bound $1s_{1/2}$ state has been  found. Since,
experimentally,  the
n$+^9$Li has a large and negative scattering length, this may 
indicate the presence of the antibound $1s_{1/2}$ state in $^{10}$Li close
to the 1n-threshold, though the presence of a  low-lying $0p_{1/2}$ state cannot be
ruled out \cite{(Boh97)}.

The GSM calculations that  explicitly consider  the antibound $1s_{1/2}$
s.p. state \cite{(Bet04),(Bet05)} were performed recently for the g.s. 
of  $^{11}$Li.  The authors  argued  that the presence of an
antibound state was important  for
the formation of a neutron  halo. They have also noted that
the bound state wave function of $^{11}$Li  could  be expanded 
in terms of the
real-energy,  non-resonant $\ell$=0  continuum, i.e., 
without  explicit inclusion of the antibound state. Moreover, a destructive
interference between the $1s_{1/2}$ antibound s.p. state and the
associated complex-energy,  non-resonant  $s_{1/2}$ background  was
 noticed. 

This work addresses the  question of the $\ell$=0 virtual state
for the description  of a neutron halo
by studying several
complementary Berggren basis expansions and performing  GSM
calculations for $^{11}$Li in different s.p. bases. In
Sec.~\ref{one_body}, the one-body Berggren completeness relations with
and without the $s_{1/2}$ antibound state are compared in the $\ell$=0
channel. Section~\ref{two_body} describes  the g.s. of $^{11}$Li in
the schematic two-particle model  using different 
s.p.~Berggren ensembles, with and without  
the explicit inclusion of the antibound  s.p.~state
$1s_{1/2}$. The comparison of
convergence properties of the corresponding GSM calculations  with the number of
$s_{1/2}$ scattering states  is made for
the g.s. energy of $^{11}$Li. A summary of results is
given in Sec.~\ref{conclusion}.

\section{Completeness relations involving antibound $s_{1/2}$ states} \label{one_body}

\begin{figure}[htbp]
\begin{center}
\begin{tabular}{c} 
\includegraphics[height=9cm]{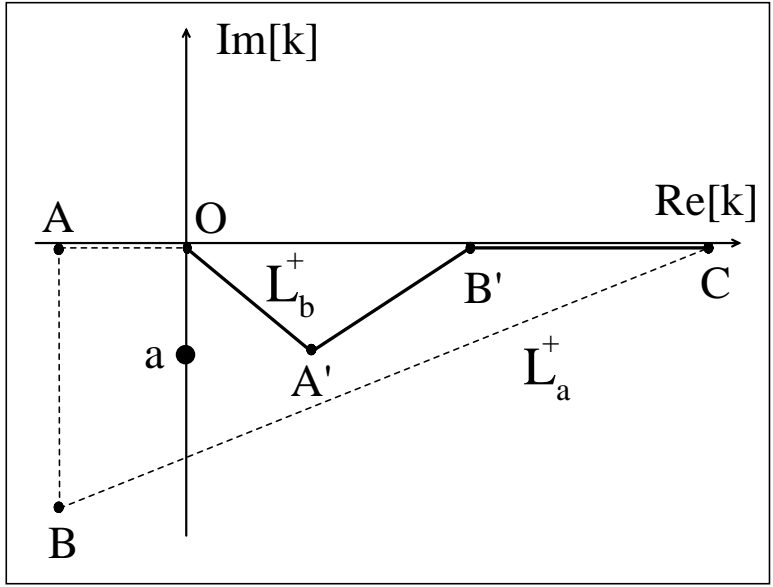}
\end{tabular}
\end{center}
\caption{Contours in the complex-$k$ plane used in the Berggren
completeness relations for the $s_{1/2}$ (Sec.~\ref{one_body} and
Sec.~\ref{two_body})  and $p_{1/2}$  (Sec.~\ref{two_body})  partial
waves. The $L^+_a$ contour (OABC) is only used in the $s_{1/2}$ channel; it
allows wielding the  antibound state  $1s_{1/2}$ (marked as `a'). The $L^+_b$ contour (OA'B'C) is employed
for $s_{1/2}$ and $p_{1/2}$ channels and permits  expansions of
bound and resonant states only.}
\label{L_plus_contours}
\end{figure}

One-body Berggren completeness relations with bound and resonant states
have already been extensively studied for neutrons 
\cite{(Ber68),(Ber93),(BerLin93),(Lin93),(Mic03)} and protons \cite{(Mic04v1),(Akram)}.
The standard Berggren completeness relation consists of a
discrete sum over bound and resonant states, and an integral over
non-resonant scattering states from the contour $L_b^+$ (see
Fig.~\ref{L_plus_contours}):
\begin{eqnarray}
\sum_{n \in (b,d)} |u_n \rangle \langle u_n| + \int_{L_{b}^+} |u(k) 
\rangle \langle u(k)| \; dk  = 1, \label{no_anti_rel}
\end{eqnarray}
where a discrete sum runs over all bound $(b)$ and
resonant decaying $(d)$ states lying above the complex contour $L_b^+$.
The 
continuous part takes into account the non-resonant scattering states
lying on the contour. 
 In the
particular case of the $\ell$=0 neutron partial wave, there are no $s_{1/2}$
resonances due to the absence of both Coulomb and centrifugal barriers.
Consequently,  a real-$k$
contour would have been sufficient to describe the  $s_{1/2}$ neutron
channel. However, to investigate the convergence of imaginary part
components of the expanded wave functions, we take the complex-$k$
contour $L_{b}^+$
close to the real-$k$ axis.

In this work, we shall perform detailed studies of the Berggren  expansion
in a more general case when an
$\ell$=0 antibound s.p. state is included 
in the basis.
In this case, the Berggren completeness relation takes the  form
\cite{(Lin93),(Bet04)}:
\begin{eqnarray}
\sum_{n \in (a,b,d)} |u_n \rangle \langle u_n| + \int_{L_a^+} |u(k) 
\rangle \langle u(k)| \; dk = 1, \label{anti_rel}
\end{eqnarray}
where the sum also includes  antibound $(a)$ states 
lying above the complex contour $L_a^+$  (see Fig.~\ref{L_plus_contours}). 
It is to be noted that, as discussed in Ref.~\cite{(Lin93)},
the contour $L_a^+$ is obtained by deforming continuously the  contour
$L_b^+$ placed in the fourth quadrant of the complex-$k$ plane 
so that it encompasses the 
antibound states of interest. In the  unlikely situation that
 bound states of energies higher than 
antibound states $(a)$ are present, they must be excluded from the
sum (\ref{anti_rel}).

Bound states can be expanded with either of the completeness relations
 (\ref{no_anti_rel}) or (\ref{anti_rel}) , whereas antibound states can
only be expressed through the more general completeness relation
(\ref{anti_rel}). Indeed, only those poles of the  $S$-matrix which are
situated above the contour can be expanded using the Berggren basis
\cite{(Ber68),(Lin93)}. The contour in the non-resonant continuum is
usually  discretized with a finite number of points in order to
construct  the Hamiltonian matrix (see Ref.~\cite{(Mic03)} for details).

In this study, the
s.p.~states entering the completeness relations (\ref{anti_rel}) and
(\ref{no_anti_rel}) have been  generated using an
auxiliary  Woods-Saxon (WS) Hamiltonian:
\begin{eqnarray}
\hat{h} &=& \frac{\hat{p}^2}{2m} - \frac{V_0}
{1 + \exp \left( \frac{r-R_0}{d}\right)}, \label{WS_Li9}
\end{eqnarray}
where $m$ is the reduced mass of a neutron with respect to the $^9$Li
core, $d$=0.65\,fm, and $R_0$=2.7\,fm.
The depth of the WS potential, $V_0$=50.5, 52.5, and 60.5\,MeV, was adjusted 
to yield the $1s_{1/2}$ eigenstate respectively antibound at $-$0.002955
MeV, loosely bound at $-$0.0329 MeV, and well bound at $-$1.0372 MeV. The
corresponding  WS potentials are denoted as $WS_1$, $WS_2$, and $WS_3$
in the following.

In order to test  Berggren completeness relations
(\ref{anti_rel}) and  (\ref{no_anti_rel}), we expand the $1s_{1/2}$
eigenstate of a given WS potential  in the basis generated by another
 WS potential (WS$^{(0)}$) of different depth:  
\begin{eqnarray}
|1s_{1/2}\rangle = \sum_{n} c_n | u_n \rangle +  \int_{L^+} c(k) | u(k) \rangle \; dk \label{u_diag}
\end{eqnarray}
where $n$ is running either over bound and antibound $|u_n\rangle$ basis states
in (\ref{anti_rel}) or over  bound $|u_n\rangle$ basis states in
(\ref{no_anti_rel}), and $L^+$ is   $L_a^+$ or $L_b^+$, respectively.
All the 
combinations of potentials studied are listed in Table~\ref{five_cases}. The
contours $L_a^+$ and $L_b^+$ used in this section 
are defined by vertices (all in fm$^{-1}$):
$[O=(0.0,0.0);A=(-0.01,0);B=(-0.01,-i0.02);C=(3.5,0.0)]$ and
$[O=(0.0,0.0);A'=(0.1,-i0.01);B'=(1.5,0.0);C=(3.5,0.0)]$, respectively.
\begin{table}
\caption{\label{five_cases} 
Set of different WS potentials and $1s_{1/2}$ states used in numerical
tests of the Berggren completeness relation. 
The Berggren ensemble
generated by a  potential WS$^{(0)}$ (second column) consists of the $0s_{1/2}$ bound
s.p.~state, the contour in the non-resonant continuum (third column),
and - possibly -  the $1s_{1/2}$ s.p. state (fourth
column).  `No pole' denotes  a situation where  the virtual $1s_{1/2}$ state is
not included 
 in the basis. In all the cases, the expansion
 has been carried out for the loosely bound $1s_{1/2}$ s.p. state  of $WS_2$.}
\begin{ruledtabular}
\begin{tabular}{|c|c|c|c|} 
Case  & WS$^{(0)}$ &  Contour & $1s_{1/2}$ (WS$^{(0)}$)  \\ \hline
(i)   & $WS_1$      & $L_a^+$     & Antibound     \\
(ii) & $WS_1$       & $L_b^+$     & No pole        \\
(iii)  & $WS_3$     & $L_b^+$     & Well bound     
\end{tabular}
\end{ruledtabular}
\end{table}

To assess the quality of the Berggren expansion,
 we
calculate the rms deviation from the exact
$1s_{1/2}$ halo wave function  $u_{WS_2}(r)$
of $WS_2$   
 obtained by a direct integration of the
Schr{\"o}dinger equation.
The rms deviation 
  is calculated separately for
the  real and imaginary parts,
\begin{eqnarray}
\mbox{rms}(\mbox{Re}[u]) &=& \sqrt{\frac{1}{N} 
\sum_{i=1}^{N} \mbox{Re}^2[u_{WS_2}(r_i) - u_{{\rm WS}^{(0)}}(r_i)] },
    \label{RMS_Re} \\
\mbox{rms}(\mbox{Im}[u]) &=& \sqrt{\frac{1}{N} \sum_{i=1}^{N} 
\mbox{Im}^2[u_{{\rm WS}^{(0)}}(r_i)] }, \label{RMS_Im}
\end{eqnarray}
for  $N$=512 equidistant points on the real $r$-axis in the interval from 
$r$=0 to $r$=15 fm. 
In Eqs. (\ref{RMS_Re}) and (\ref{RMS_Im}) $u_{{\rm WS}^{(0)}}(r)$
is the $1s_{1/2}$ halo wave function of $WS_2$  expanded in the
basis  WS$^{(0)}$ of Table~\ref{five_cases}.

\begin{figure}[htbp]
\begin{center}
\includegraphics[width=10.00cm]{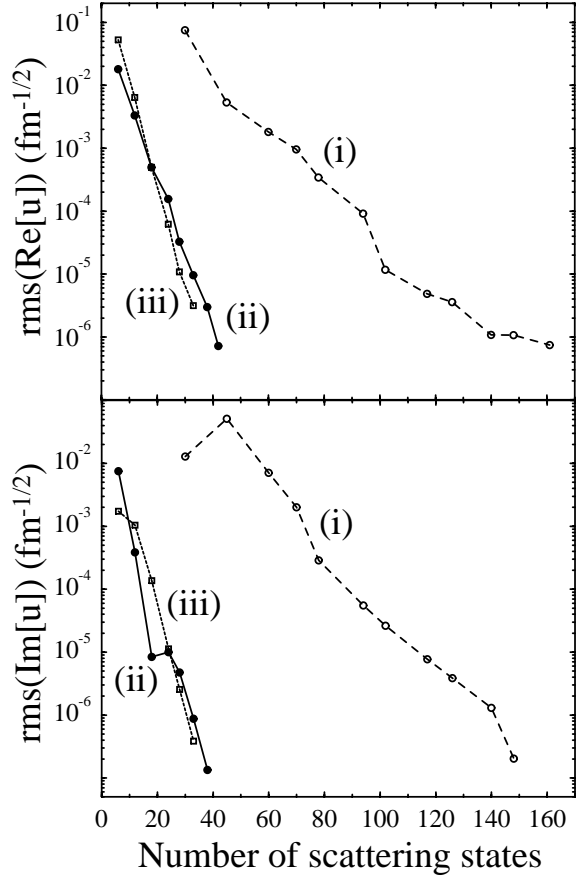} 
\end{center}
\caption{Real and imaginary parts of the rms deviations (\ref{RMS_Re}, \ref{RMS_Im})
for the $1s_{1/2}$ halo wave function expanded in different
 Berggren bases  as a function of 
the number of $s_{1/2}$ scattering wave functions on $L^+$. 
Cases (i)-(iii)  of Table
\ref{five_cases} are marked by  dashed,  solid,
and  dotted lines,  respectively.}
\label{RMS_wf}
\end{figure}

The calculated rms deviations
(\ref{RMS_Re}) and (\ref{RMS_Im})  are shown in Fig.~\ref{RMS_wf}. 
One clearly sees that the
Berggren basis containing an antibound state (case (i) in Table
\ref{five_cases}) is less efficient in expanding a loosely bound
$1s_{1/2}$ state.  The number of discretized scattering states in case
(i) must be two-to-four times bigger than that in cases (ii) and (iii)
in order to attain the same precision for the real part of the wave
function. For the imaginary part, the difference is even more pronounced.

Without an antibound state in the basis, 40 to 50 non-resonant
scattering states are enough to obtain  the precision of order $10^{-6}$
for the calculated $s_{1/2}$ wave function,  whereas 150 non-resonant
scattering states are necessary to reach the same precision with this
state included. In cases (ii) and (iii), one finds similar rms
deviations, because
the  $s_{1/2}$ basis wave functions are in both cases 
either bound or close to the real (positive) $k$-axis; hence their
contributions add up constructively. On the contrary,  in case (i),
the virtual state with exponentially increasing wave function
interferes destructively with
non-resonant scattering states  in order to produce the halo state.

\begin{figure}[htbp]
\begin{center}
\includegraphics[width=10.0cm]{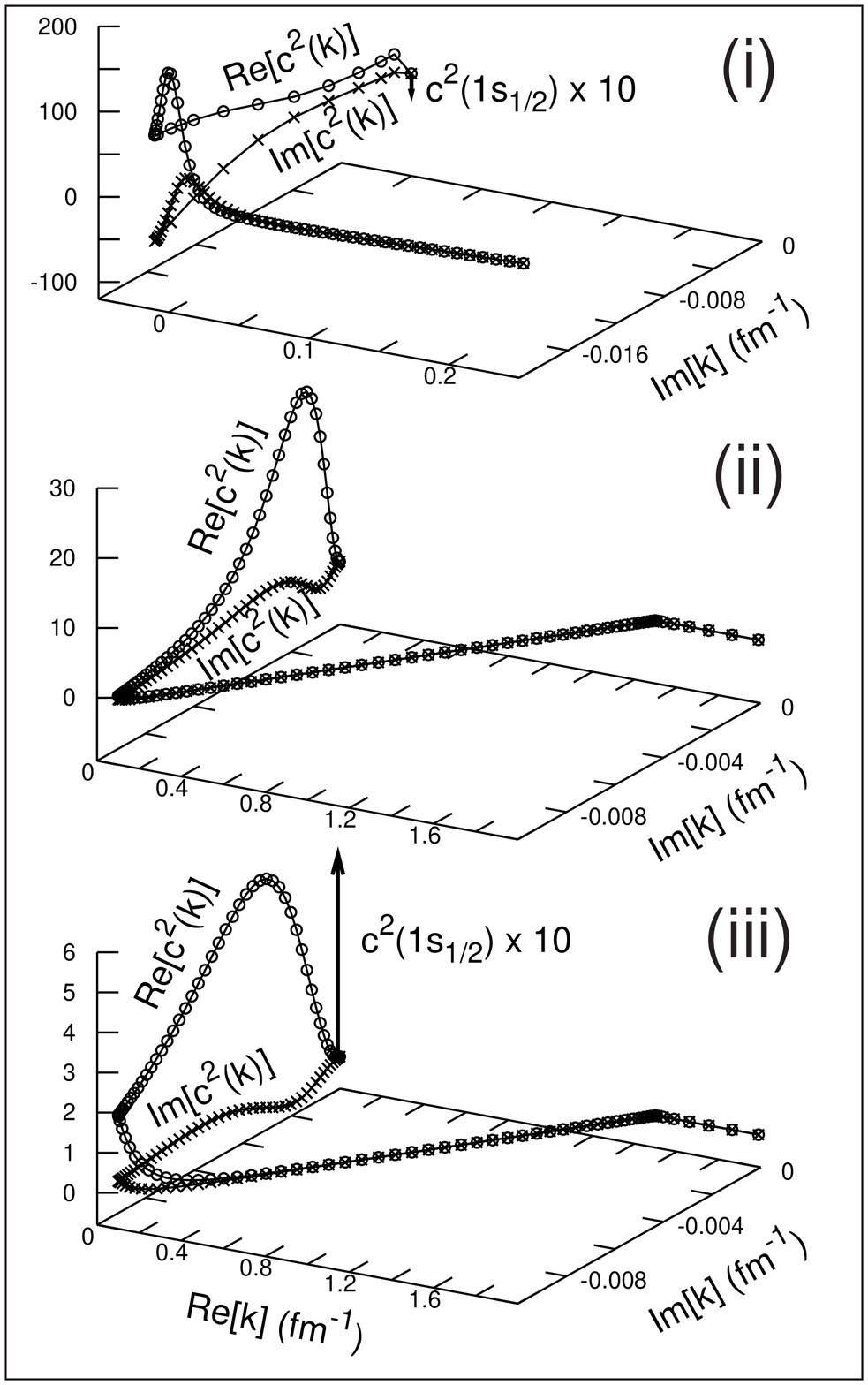} 
\end{center}
\caption{Real (circles) and imaginary (crosses)  parts of the squared
expansion amplitudes $c_n^2$ and $c^2(k)$ (\ref{u_diag}) 
corresponding to  the
Berggren bases (i-iii) of Table~\ref{five_cases}
in the complex $k$-plane for
the  $1s_{1/2}$ halo wave function.
The amplitude $c^2_{0s_{1/2}}$ is practically negligible. 
In  cases (i) and (iii), the amplitude
 $c_{1s_{1/2}}^2$ (magnified by a factor of 10)
is represented by an arrow placed arbitrarily at $k$ = 0.} 
\label{squared_amplis}
\end{figure}

To illustrate this effect, Fig.~\ref{squared_amplis} shows
the distribution of squared amplitudes  $c_n^2$ and $c^2(k)$
(\ref{u_diag}).
By construction, the sum $\displaystyle \sum_n c_n^2+\int c^2(k) dk$ is
always normalized  to one. 
As the contribution from the $0s_{1/2}$ basis state is
practically negligible,
the set of $c_n^2$ reduces to the single real
amplitude $c^2_{1s_{1/2}}$.
In  case (i),
the non-resonant amplitudes $c^2(k)$ are much larger 
than their 
antibound counterpart  $c_{1s_{1/2}}^2$, whereas in  case (iii) the
bound $1s_{1/2}$ pole carries half of the halo wave function. 
Case (ii) lies between these two extremes, as the halo
state is  built entirely from the non-resonant
continuum but in an essentially coherent way  (cf. Refs.~\cite{(Bet04),(Mig71)}). 

The results displayed in Fig.~\ref{squared_amplis} demonstrate that the
inclusion of the antibound pole in the basis  enormously enhances the
role of the non-resonant continuum which has to efface the incorrect
asymptotics of an antibound pole in order to create a bound state with
decaying asymptotics.  This behavior is opposite to what is found when
including  a narrow resonant state in the Berggren ensemble. Such a
state always concentrates a fairly large part of the expanded wave
function \cite{(Mic03),(Mic04v1)}.

\section{Contribution of antibound states to the two-body halo states: example of $^{11}$$\mbox{Li}$} \label{two_body}

In order to assess the influence of antibound states on many-body halo states, we
shall employ a schematic model  \cite{(Bet04),(Bet05)} of two valence particles,
coupled to $J^\pi$=$0^+$, moving outside the inert core. In particular,
we shall view the g.s. of  $^{11}$Li in terms of
two neutrons in $s$ and $p$ angular momentum states 
 coupled to the $^{9}$Li core. Our aim is not 
to exactly reproduce the structure of 
$^{11}$Li but to learn  how antibound s.p. states enter 
a two-body halo wave function. 

The nuclear Hamiltonian is given by a WS potential representing 
a $^9$Li
core, to which a Surface Gaussian Interaction (SGI)
\cite{(Mic04v1)} is added, modeling the residual interaction
between the two valence nucleons:
\begin{eqnarray}
\hat{H} &=& \sum_{i} \left[ \frac{\hat{p}_i^2}{2m} - V_0 \cdot f(r_i) 
 - 4 \; V_{so} \left( \vec{l_i} \cdot \vec{s_i} \right) \frac{1}{r_i} \left| \frac{df(r_i)}{dr} \right|
 \right] \nonumber \\
  &+& \sum_{i < j} V_{\rm SGI} \cdot \exp \left( - 
  \left[ \frac{\vec{r_i} - \vec{r_j}}{\mu} \right]^2 
  \right) \cdot \delta(|\vec{r_i}| + |\vec{r_j}| - 2 R_{\rm SGI}).
\label{H}
\end{eqnarray}
Here, we employ the set of WS parameters  $WS_1$ of  
Sec.~\ref{one_body}, and $V_{so}$=21.915\,MeV is the strength of the
spin-orbit potential. For these parameters, the $1s_{1/2}$ state is
antibound and the $0p_{1/2}$ state is a resonance with the energy
$E$=0.24\,MeV and the width $\Gamma$=118 keV,  in  fair agreement  with
experimental data \cite{(Tho94),(Boh97)}.  The range of the residual SGI
interaction  is $\mu$=1\, fm,   $R_{\rm SGI}$=4.4 fm, and 
$V_{\rm SGI}$=$-1196$ MeV fm$^3$. These parameters
have been  adjusted to
reproduce the binding energy of $^{11}$Li  with respect to $^9$Li
($E_B$=--0.295\,MeV),  and a nearly 50\% weight of the $s^2$ component,  as
suggested by the data \cite{(Tho94)}.

The valence space consists of $s_{1/2}$, $p_{3/2}$, and $p_{1/2}$ neutron
wave functions, from which the $0s_{1/2}$ and $0p_{3/2}$ bound states have
been removed as they are part of the $^9$Li core. The $p_{3/2}$ s.p. space
contains the $0p_{3/2}$ bound state pole and the real-energy
$p_{3/2}$ contour with $k_{\rm max}$=2 fm$^{-1}$. The $p_{1/2}$ space
consists of the $0p_{1/2}$ resonant pole and the $p_{1/2}$ contour of an
$L_b^+$ type (see Fig.~\ref{L_plus_contours}) defined by the vertices
(all in fm$^{-1}$): 
$[O=(0.0,0.0); A'=(0.1,-i0.1); B'=(1.0,0.0); C=(2.0,0.0)]$. The $s_{1/2}$
space  contains  either a  scattering component  ($L_b^+$) or the $1s_{1/2}$
antibound pole to which the $s_{1/2}$ contour of an $L_a^+$-type is added. 
The contours $L_a^+$ and $L_b^+$ in the
 $s_{1/2}$ channel are defined by the
vertices (all in fm$^{-1}$): $[O=(0.0,0.0); A=(-0.01,0); B=(-0.01,-i0.04); C=(2.0,0.0)]$ and
$[O=(0.0,0.0); A'=(0.1,-i0.01); B'=(1.0,0.0); C=(2.0,0.0)]$, respectively.
The
$p_{3/2}$ and $p_{1/2}$ contours are discretized with 30 and 32 points,
respectively. Each point represents one shell in GSM calculations. For
this level of contour discretization and the momentum cut-off,
the theoretical error on energies  is about 1 keV for the real part and 0.01 keV for
the imaginary part. 

\begin{table}
\caption{\label{E_cv_table} Energy $E$ (in MeV)
 and width $\Gamma$ (in keV) of the $^{11}$Li g.s. as a function of 
 the number $N_{s_{1/2}}$ of the non-resonant scattering shells in the 
 discretized  $s_{1/2}$ continuum. The values of $E/\Gamma[L_a^+]$ (second and fourth columns) 
 are calculated with the 
$s_{1/2}$ space consisting of the $1s_{1/2}$ antibound state 
 and the associated $L_a^+$ $s_{1/2}$ contour. The values of
 $E/\Gamma[L_b^+]$ (third and fifth columns) are obtained with the $s_{1/2}$
 space consisting of the $L_b^+$ $s_{1/2}$ contour only. 
See Fig.~\ref{L_plus_contours} for the definition of different contours.}
\begin{ruledtabular}
\begin{tabular}{|c|c|c|c|c|} 
$N_{s_{1/2}}$   & $E[L_a^+]$  & $E[L_b^+]$  & $\Gamma[L_a^+]$  & $\Gamma[L_b^+]$ \\ \hline
10              & --0.314          & --0.291          &  65.274               & --3.644               \\
20              & --0.292          & --0.295          &   2.307               &   0.025               \\
30              & --0.294          & --0.295          &   0.876               & --0.003               \\
40              & --0.294          & --0.295          & --0.425               & --0.007               \\
50              & --0.295          & --0.295          &   0.075               & --0.009               \\
60              & --0.295          & --0.295          & --0.005               & --0.009      
\end{tabular}
\end{ruledtabular}
\end{table}

The convergence of the $^{11}$Li g.s. energy as a function of the number of
$s_{1/2}$ non-resonant scattering shells is shown in
Table~\ref{E_cv_table}. The number of $s_{1/2}$ shells on each segment
of the contours $L_a^+$ and $L_b^+$ is chosen so as to minimize a
spurious width of the g.s. It is seen that the energy converges  much faster
when the  antibound state is not present in the s.p. basis. In this case, the calculated
g.s. energy of $^{11}$Li attains a precision of 0.1 keV with 20 shells
included, whereas as many as 50 shells are necessary to
obtain the same level of precision if  the antibound
pole is present.

It is  instructive to inspect  the g.s. wave function of $^{11}$Li 
expressed in different Berggren bases. Table \ref{GS_config} compares the
GSM results obtained in the Berggren basis including the $1s_{1/2}$
antibound pole and $s_{1/2}$ non-resonant scattering states from the
$L_a^+$ contour (case (i)  of Table \ref{five_cases}) with those
obtained using the contour $L_b^+$ for the $s_{1/2}$ part (case (ii)).
(The $p_{3/2}$ and $p_{1/2}$ spaces are defined as  above.) To determine
precisely the valence neutron configurations, 60 shells were taken along
each of the scattering contours: $s_{1/2}$, $p_{1/2}$, and $p_{3/2}$. 
The expansion  amplitudes of the $p$ components are identical in both
cases and tote up to about 49\%. As expected, important
cancellations appear in  case (i). Here, about 50\% of the sum of
squared amplitudes for  configurations with two neutrons in $s_{1/2}$
non-resonant scattering states is cancelled by 
contributions from configurations  involving  one neutron in
a  scattering state and  another one  in the $1s_{1/2}$ antibound
state. The antibound-antibound  configuration $(1s_{1/2})^2$ contributes
only   $\sim$10\%. In   case (ii), however,  all scattering two-neutron
configurations add up coherently. 

\begin{table}
\caption{\label{GS_config} Real and imaginary parts of  squared
amplitudes of the GSM configurations involving neutrons in 
$s_{1/2}$ orbits in the g.s. of $^{11}$Li. The GSM calculations were
performed using two Berggren bases: (a) including the $1s_{1/2}$
antibound pole state and the  non-resonant scattering states $s_{1/2}$ along
 $L_a^+$  (second and fourth columns), and (b)
including  
the  non-resonant scattering states  $s_{1/2}$ along $L_b^+$  
(third and fifth columns). For the definition of contours, see
Fig.~\ref{L_plus_contours}.
}
\begin{center}
\begin{ruledtabular}
\begin{tabular}{|c|c|c|c|c|} 
Configuration          & Re$[c^2]\{L_a^+\}$ & Re$[c^2]\{L_b^+\}$ &  Im$[c^2]\{L_a^+\}$  &  Im$[c^2]\{L_b^+\}$   \\ \hline
$(1s_{1/2})^2$         &   0.0990             &     --                  & --9.6033$\cdot 10^{-6}$     & --                      \\ 
($1s_{1/2}$ $s_{1/2}$) & --0.5887             &     --                  &   2.3369$\cdot 10^{-5}$     & --                      \\
$(s_{1/2})^2$          &   1.0034             &   0.5137                & --8.0720$\cdot 10^{-6}$     & --1.4650$\cdot 10^{-5}$ \\
\end{tabular}
\end{ruledtabular}
\end{center}
\end{table}

\section{Conclusions} \label{conclusion}

The GSM has proven to be a powerful approach for the microscopic
description of loosely bound and resonant states. The fact that
the underlying Berggren basis directly incorporates one-particle continuum
and the proper treatment  of many-body
correlations through configuration mixing  makes
it a perfect tool for  a theoretical description of loosely bound many-body
states such as  nuclear halos. 

In this study, we investigated the importance of including the virtual
$\ell$=0 state in the s.p.  basis for a modeling 
of one-body and two-body
neutron halos. This question has experimental
relevance: the data seem to  suggest that
the presence of the $s_{1/2}$ antibound state  in $^{10}$Li is correlated
with the appearance of the two-neutron halo  in $^{11}$Li. 

Our calculation suggests that  neither conceptual nor numerical gain is
achieved if the antibound states are directly considered in the Berggren
ensemble. In both one- and two-particle cases, it is always more
advantageous  to use a s.p. basis which contains   non-resonant scattering 
states  and, possibly, a bound $s_{1/2}$ pole. As the halo wave function
has a decaying character at large distances due to the exponentially
increasing asymptotics of the virtual state, its inclusion  in the basis
always induces strong negative interference with the scattering states.
Therefore, adding antibound states to the basis is not beneficial, as
more discretized scattering $s_{1/2}$ states are  necessary to reach a
required precision without providing any new information about the
many-body wave function.

The use of antibound states in a Berggren basis can only be justified  if
the many-body state has a virtual character. This is certainly not the
case for bound states, including   halos. For instance, the  g.s.
halo  in $^{11}$Li is  well described by using  the Berggren basis
solely involving the  $s_{1/2}$ non-resonant scattering continuum  and
choosing  the integration contour close to the real-$k$ axis. While the
presence of an antibound state does require an increased density of
discretized states around $k$=0, this can be handled efficiently by
employing  the density matrix renormalization group approach
\cite{(Rot06a)}. The resulting procedure   yields much better numerical
precision by suppressing the  cancellations and leaving all the physical
properties unchanged. Hence, the use of antibound states in GSM 
becomes questionable, as it worsens  spurious effects due to continuum
discretization.

\begin{acknowledgments}
This work was supported in part by the U.S.~Department of Energy
under Contracts Nos.~DE-FG02-96ER40963 (University of Tennessee),
DE-AC05-00OR22725 with UT-Battelle, LLC (Oak Ridge National
Laboratory) and DE-FG05-87ER40361 (Joint Institute for Heavy Ion
Research).
\end{acknowledgments}

\end{document}